\documentclass[5p]{elsarticle}

\usepackage{hyperref}
\hypersetup{
    colorlinks  = false,
    pdfborder   = {0 0 0},
    pdftitle    = {Efficient determination of HPGe gamma-ray efficiencies at high energies with ready-to-use simulation software},
    pdfauthor   = {Jan Mayer},
}

\usepackage[T1]{fontenc}
\usepackage[utf8]{inputenc}
\usepackage{newunicodechar}
\newunicodechar{γ}{\ensuremath\gamma}
\newunicodechar{Δ}{\ensuremath\Delta}
\newunicodechar{Θ}{\ensuremath\theta}
\newunicodechar{φ}{\ensuremath\phi}
\newunicodechar{α}{\ensuremath\alpha}

\usepackage{siunitx}
\usepackage{booktabs}
\usepackage{tikz}
\usetikzlibrary{arrows.meta}
\usepackage{url}
\usepackage[capitalise]{cleveref}
\usepackage{isotope}
\usepackage{csquotes}
\usepackage{microtype}

\journal{Nuclear Instruments and Methods in Physics Research Section A}
\begin{document}

\begin{frontmatter}

\title{Efficient determination of HPGe γ-ray efficiencies at high energies with ready-to-use simulation software}

\author{Jan Mayer\corref{corres}}
\ead{jan.mayer@ikp.uni-koeln.de}
\author{Elena Hoemann}
\author{Markus Müllenmeister}
\author{Philipp Scholz\fnref{nd}}
\author{Andreas Zilges}
\address{University of Cologne, Institute for Nuclear Physics, Zülpicher Strasse 77, 50937 Köln, Germany}

\cortext[corres]{Corresponding author}
\fntext[nd]{Present Address: University of Notre Dame, Department of Physics, Indiana 46556-5670, USA}

\begin{abstract}
The full-energy-peak efficiency of HPGe detectors at γ-ray energies around \SI{10}{\MeV} is not easily accessible with experimental methods.
Monte-Carlo simulations with \textsc{Geant4} can provide these efficiencies.
G4Horus is a ready-to-use \textsc{Geant4} application for the HORUS HPGe-detector array.
Users can configure the modular parts to match their experiment with minimal knowledge of the simulation software and limited time commitment.
In our case, knowing and implementing the geometry with high precision is the biggest challenge.
To implement the different target chambers, we transform the existing CAD models to \textsc{Geant4} geometry with \textsc{CADMesh}.
We also found a large discrepancy between experimental and simulated efficiency for some older HPGe detectors, which could be remedied by introducing a large dead region around the inner core.
This project is open source and available from \url{https://github.com/janmayer/G4Horus} \cite{jan_mayer_2020_3692475}.
We invite everyone to adapt the project or adopt parts of the code for other projects.

\hspace*{0.2mm}

\noindent Published in Nuclear Instruments and Methods in Physics Research Section A 972, 164102 (2020).

\noindent \href{https://doi.org/10.1016/j.nima.2020.164102}{DOI: 10.1016/j.nima.2020.164102} © 2020. This manuscript version is made available under the CC-BY-NC-ND 4.0 license \url{http://creativecommons.org/licenses/by-nc-nd/4.0/}.
\end{abstract}


\end{frontmatter}

\section{Introduction}
In high-resolution γ-ray spectroscopy, efficiency is a significant attribute.
While a large detection efficiency is beneficial for data collection, it is the precise value (and its energy dependency) which is crucial for data analysis.

The γ-ray energy range of interest heavily depends on the experiment.
Most γ-rays observed in nuclear physics stem from transitions between excited states of nuclei.
These commonly have energies between \SI{100}{\keV} and \SI{3}{\MeV}, and thus γ-ray spectroscopy is often performed in this energy region.

Several research areas have come into focus which require γ-ray detection at energies around \SI{10}{\MeV} and higher, for example studies of the Pygmy Dipole Resonance (PDR) and radiative capture reactions for nuclear astrophysics.
For the PDR, the decay behavior of $J^\pi=1^-$ states at energies below the neutron separation energy is studied \cite{Savran2013}.
This includes direct decays to the ground state, i.e., γ-ray transitions around \SIrange{5}{10}{\MeV}.
For radiative capture reactions, direct transitions from the entry state at the sum of center-of-mass energy and Q-value to the ground state must be investigated \cite{Netterdon2015}.
This translates to γ-ray energies up to \SI{15}{MeV}.

The higher the γ-ray energy, the harder becomes a reliable experimental determination of the efficiency.
Standard sources provide calibration up to \SI{3.6}{\MeV} only.
From thereon, fewer and more complex methods can be used, see \cref{c:excal}.
Our areas of research require precise efficiency calibration at energies hardly accessible experimentally.

Simulations can address this need for fast, easy, and reliable calibration at any γ-ray energy.
Interactions of γ-rays with matter are known well enough; and given geometries and materials, Monte-Carlo simulations with particle transport codes like \textsc{Geant4} \cite{Agostinelli2003} can provide full-energy-peak (FEP), single-escape-peak (SEP), double-escape-peak (DEP), and coincidence efficiencies.
\textsc{Geant4} provides a simulation framework, but no ready-to-use executable -- one must implement each specific setup.

G4Horus provides a ready-to-use \textsc{Geant4} based application for simulating the efficiency of γ-ray detectors.
It is used at the Institute for Nuclear Physics, University of Cologne, to simulate the efficiency of the HPGe-detector array HORUS, see \cref{c:horus}.
It provides everything required to simulate the efficiency, that includes especially detector and target chamber geometries and a predefined workflow that requires minimal knowledge and effort from the user.

\subsection{\texorpdfstring{γ}{Gamma}-ray spectroscopy with HORUS}\label{c:horus}
Located at the \SI{10}{MV} FN-Tandem accelerator at the Institute for Nuclear Physics, University of Cologne, the γ-ray spectrometer HORUS (High-efficiency Observatory foR Unique Spectroscopy) is used to investigate the structure of nuclei and measure cross sections to answer questions in nuclear astrophysics.

It consists of up to 14 HPGe detectors, six of which are equipped with active anti-Comp\-ton BGO shields \cite{Netterdon2014a}.
Signals from the detectors are processed by XIA's Digital Gamma Finder 4C Rev.\,F, which allows for acquisition of so-called \emph{listmode} data, where coincident hits in different detectors can be correlated \cite{Pickstone2012}.
For example, γγ coincidences can be used to investigate quadrupole and octupole states \cite{Pascu2015} or low-spin structures \cite{Fransen2004}.

Passivated Implanted Planar Silicon (PIPS) particle detectors can be added with the SONIC detector chamber \cite{Pickstone2017}.
They are used in coincidence with the HPGe detectors to select events with a specific excitation energy, which eliminates other unwanted feeding transitions.
The resulting spectra are used for lifetime measurements with the DSAM technique \cite{Hennig2015} or to investigate the Pygmy Dipole Resonance \cite{Pickstone2015}.

In addition, high energetic γ-rays, which are emitted after capture of protons or α-particles, can be used to determine total and partial cross sections for nuclear astrophysics \cite{Netterdon2014a, Mayer2016}.

HORUS has no default, fixed configuration. For every experiment, the detectors and target chambers are optimized to match the experimental requirements.

\subsection{Experimental efficiency calibration}\label{c:excal}
The full-energy-peak efficiency can be determined experimentally using standardized calibration sources and known reactions.

Standard sources of not-too-short lived radioactive isotopes provide easily accessible calibration points up to \SI{3.6}{\MeV} and thus are commonly used for both energy and efficiency calibration.
Sources with known activity made from, e.g., \isotope[152]{Eu} and \isotope[226]{Ra}, are excellent for the γ-ray-energy range up to \SI{3}{\MeV}.
As their half-lifes span decades, they only need to be procured once.
\isotope[56]{Co} emits usable γ-rays up to \SI{3.6}{\MeV}.
Due to its half-life of \SI{77}{\day}, sources need to be re-activated about every year via the (p,n) reaction on an enriched \isotope[56]{Fe} target.

More exotic isotopes can extend the coverage up to \SI{5}{\MeV}.
The energy range covered by the 69 nuclides included in the IAEA xgamma standard \cite{iaea-xgamma} ends at \SI{4.8}{\MeV} with the isotope \isotope[66]{Ga}.
The Decay Data Evaluation Project (DDEP) \cite{DDEP} lists several more exotic nuclei.
Here, the highest transition at \SI{5}{\MeV} also stems from \isotope[66]{Ga}.
With an almost negligible intensity of \SI{0.00124\pm0.00018}{\percent}, it is, however, not well suited for calibration purposes.
While the energy range covered by \isotope[66]{Ga} is expedient, the short half-life of \SI{9.5}{\hour} is not and requires the source to be produced anew for each project -- increasing the already high workload of the main experiment.

Decay measurements of short-lived isotopes in target position can extend the energy range up to \SI{11}{\MeV}.
The decay of \isotope[24]{Al} with a half-life of \SI{2}{\s}, created by pulsed activation of \isotope[24]{Mg}, is a feasible way to obtain calibration lines up to \SI{10}{\MeV} \cite{Wilhelm1996, Pickstone2017}.
Neither the IAEA nor the DDEP currently include \isotope[24]{Al} in their list of recommended nuclides, thus there can be doubts on the accuracy of the existing decay intensity data.
This method is even more involved than the methods mentioned before, as a pulsing device must be set up at the accelerator injection and linked to the data acquisition.
In addition, this method releases neutrons close to the HPGe detectors, which might be damaged.

Direct γ-ray emissions from capture reactions can also be used for efficiency calibration.
Emissions from neutron capture reactions, mostly \isotope[14]{N}(n,γ)\isotope[15]{N}, have been used successfully \cite{Molnar2002, Belgya2008, MIYAZAKI2008}.
As this method requires neutrons, which are neither trivial to procure nor healthy for HPGe detectors, we have made no efforts to introduce this method at HORUS.

We have previously used direct γ-ray emissions from the proton capture resonance of \isotope[27]{Al}(p,γ)\isotope[28]{Si} at $E_p = \SI{3674.4}{\keV}$ \cite{Netterdon2014a}.
As the measurements take about a day, the intensity uncertainties are high, and angular distributions must be corrected for, we no longer perform these regularly.
The \isotope[27]{Al}(p,γ)\isotope[28]{Si} reaction has many resonances, however only few have been measured extensively, e.g., at $E_p = \SI{992}{\keV}$ \cite{Scott1975}.
There are also several resonant proton capture reactions on other light isotopes, e.g., on \isotope[23]{Na}, \isotope[39]{K}, \isotope[11]{B}, \isotope[7]{Li} \cite{Elekes2003,Zijderhand1990,Ciemaa2009}, and \isotope[13]{C} \cite{Kiener2004}.
Unfortunately, these comparatively low-lying resonances are hard to reach with the high-energy FN-Tandem accelerator -- they might be perfectly accessible for other groups.

Alternatively, given enough calibration points, extrapolation using fitted functions can be used. This process can produce diverging results for large distances from the highest calibration point and choice of fit function \cite{Molnar2002}, but is reasonably accurate otherwise and low-effort.

To Summarize: A thorough γ-ray efficiency calibration uses up more time and effort the higher the γ-ray energy of interest.

\section{Purpose}\label{c:purpose}
We developed G4Horus to provide several services to support experiments at HORUS.
The goals in order of importance are:

1) Provide accurate full-energy-peak efficiency.
The difficult access to calibration points at high energies as described in \cref{c:excal} leaves a gap which Monte-Carlo simulations can fill.
Simultaneously, they can provide the single- (SEP) and double-escape-peak (DEP) efficiency with and without active veto signal from the BGO anti-Compton shields.

2) Require minimum effort and domain-specific knowledge from the user.
\textsc{Geant4} does not offer a ready-to-use application and even to get \emph{just} the efficiency, a full implementation of all components is required.
All users should be able to use the software without having to worry about knowing \textsc{Geant4} and without spending more time than necessary.

3) Adapt to all experimental configurations.
The HORUS setup is highly configurable with many different detectors, target chambers, and other equipment.
Users should be able to reproduce their individual configuration from predefined modular parts.

4) Guide new developments.
Experimental requirements continuously change.
Simulations can help to make informed decisions for adaptations to the setup.

5) Provide coincidence and other high-level data.
With simulations, coincidence efficiencies can be checked, and the correctness of the analysis-software procedure confirmed. They can also be used to develop and test new experimental setups and analysis methods.

\section{Implementation}
Monte Carlo simulations of γ-ray detectors are well established \cite{Hardy2002, Soderstrom2011, Baccouche2012}.
For \textsc{Geant4}, the three main components geometry, physics, and actions must be implemented.

The main difficulty is summarized well in \cite{Giubrone2016}:
\enquote{The accuracy of \textsc{Geant4} simulations is heavily dependent on the modeled detector geometry. Characterizing a detector is difficult, especially if its technical characteristics are not well known.}
This especially applies to HPGe detectors, where the manufacturer often only provides the most basic information, e.g., crystal size and weight.
X-ray imaging is a non-destructive method to obtain excellent geometry data for the crystal \cite{Chuong2016}, however not the full volume of the crystal might be \emph{active} volume, see \cref{c:geocoax}.

Passive materials between the source and the detector must be implemented accurately as well.
Users of Monte-Carlo simulation software commonly manufacture the desired shapes by writing code to create, intersect, and position basic shapes.
This seems excessively complicated compared to industry standard engineering tools.
In our case, the complex shapes of the CNC-milled target chambers are difficult or even impossible to implement with standard \textsc{Geant4} tools.
Instead, we use CAD files directly, see \cref{c:chambergeo}.

\subsection{Geometry}
\subsubsection{Target chambers and CAD-based geometry}\label{c:chambergeo}
In general, geometry in \textsc{Geant4} is implemented by writing \texttt{C++} code.
Basic shapes like boxes and spheres are created, rotated, intersected, and placed manually without visual interfaces.
While this is feasible for simple volumes, more complicated structures might be drastically reduced in detail or simply skipped and not implemented at all.
Such a simplified geometry might be acceptable or even desired for faster execution in some cases.
However, investigations of, e.g., background caused by passive components, are meaningless without all physical structures placed completely and accurately.

The target chambers used at HORUS are, like most modern mechanical structures, created using Computer Aided Design (CAD) software, and then build with Computer Numerical Control (CNC) milling machines or even 3D printers.
We think that not using these CAD-files, which already exist \emph{anyway}, is a massive waste of time and effort, independent of the complexity of the models.
Even if these do not exist yet, it should be significantly faster and less error prone to re-create them with a CAD program instead of writing \texttt{C++}-\textsc{Geant4} code.

There are several concepts for creating \textsc{Geant4} compatible volumes from CAD models.
If the shape has been constructed with Constructive Solid Geometry (CSG), the underlying configuration of basic components can be translated to basic \textsc{Geant4} shapes and Boolean operations.
In principle, this is the favorable solution, as it is simple yet elegant and might offer the best performance during simulation.
If the CSG configuration is not known, it is sometimes possible to recreate it with CSG decomposition \cite{Lu2017a}.
Complex volumes can also be converted to a tessellated shape, where the surface is represented by a triangle mesh, called \texttt{G4TessellatedSolid} in \textsc{Geant4} \cite{Poole2012}.
Alternatively, the whole volume can be split into many tiny tetrahedrons (\texttt{G4Tet}) using a Delaunay-based algorithm \cite{Si2015}.
A hybrid approach, that is building a simple structure with CGS and adding complex details with tessellated meshes, is also conceivable.
Converted shapes can be stored in the \texttt{GDML} (Geometry Description Markup Language) format.

The idea of using these CAD files in \textsc{Geant4} is not new, but there is no widely adopted solution.
A conversion can either be performed with plugins in the CAD program, a standalone software, or as a dependency in the \textsc{Geant4} application itself.
For example, plugins have once been developed for \emph{FreeCAD} \cite{FreeCADGDML, Pinto2019} and \emph{CATIA} \cite{Belogurov2011}.
Notable standalone projects are \emph{cad-to-geant4-converter} \cite{Tykhonov2015}, \emph{STEP-to-ROOT} \cite{Stockmanns2012a}, \emph{SW2GDMLconverter} \cite{Vuosalo2016}, and \emph{McCad-Salome} \cite{Lu2017a}.
Some projects seem to be abandoned, having received their last update several years ago.

We had success with \emph{CADMesh} \cite{Poole2012a} to integrate our geometry.
The CADMesh software package supports creating tessellated and tetrahedral meshes in \textsc{Geant4} at runtime, which enables fast iteration and a flexible geometry selection.

The sequence of operations is as follows:
We receive the original geometry as STEP file from the mechanical workshop, which includes every detail as its own object.

\begin{figure}
\centering
\includegraphics[width=0.49\columnwidth, height=0.495\columnwidth]{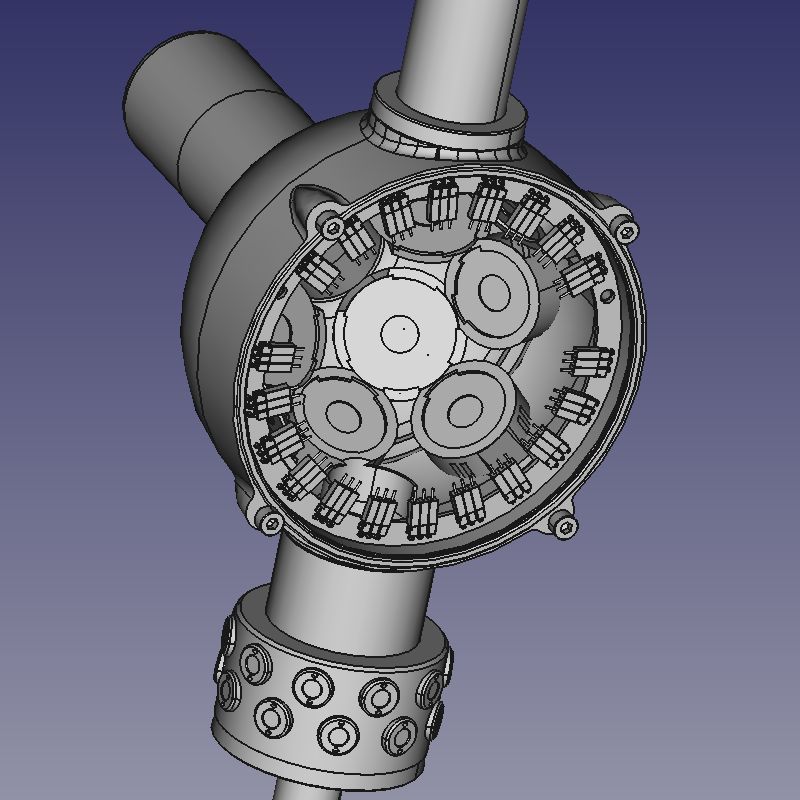}
\includegraphics[width=0.49\columnwidth, height=0.495\columnwidth]{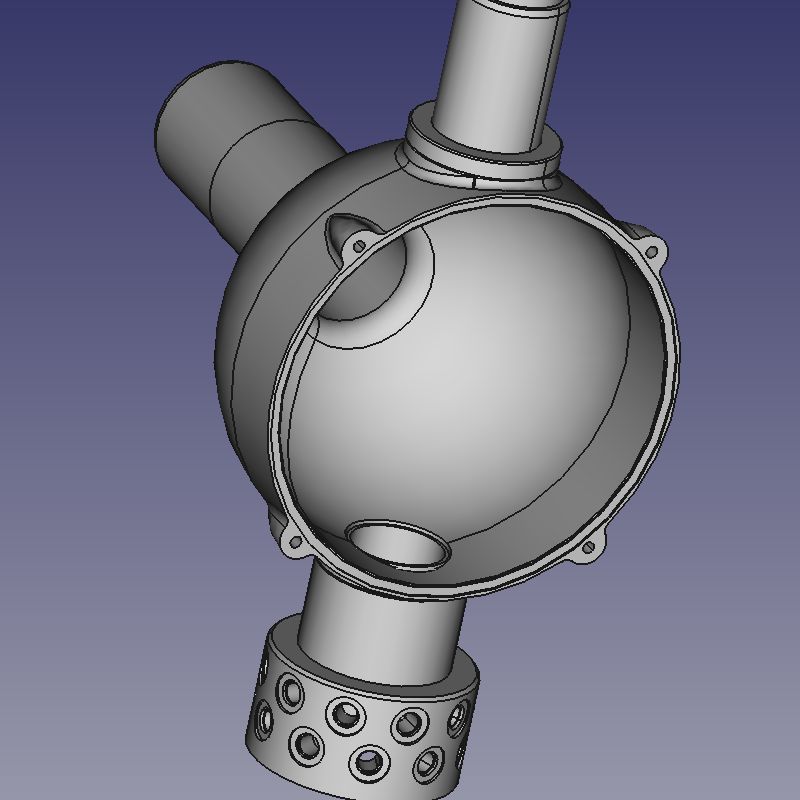}
\includegraphics[width=0.49\columnwidth, height=0.495\columnwidth]{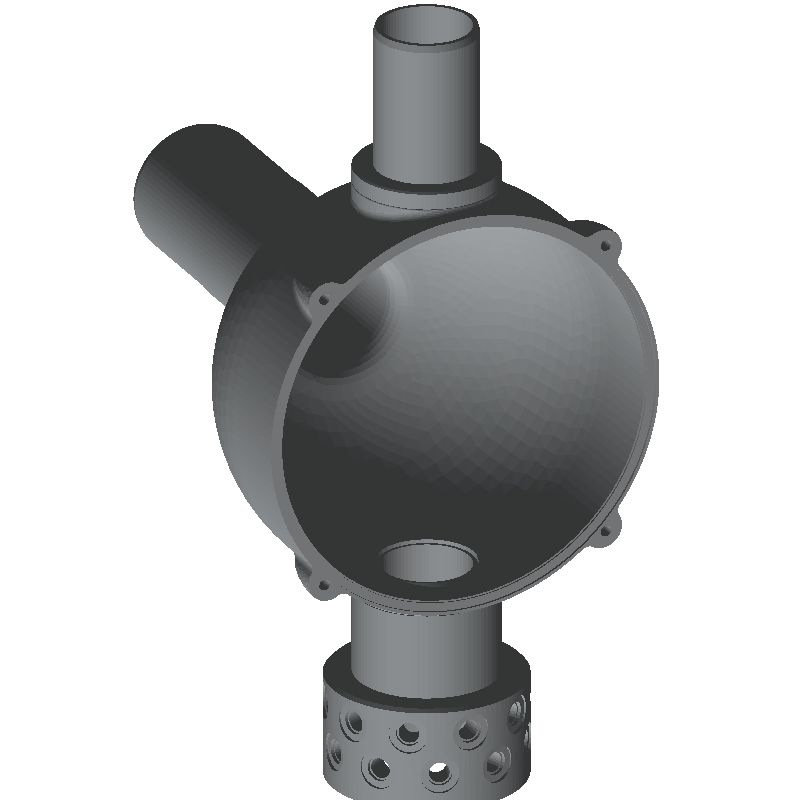}
\includegraphics[width=0.49\columnwidth, height=0.495\columnwidth]{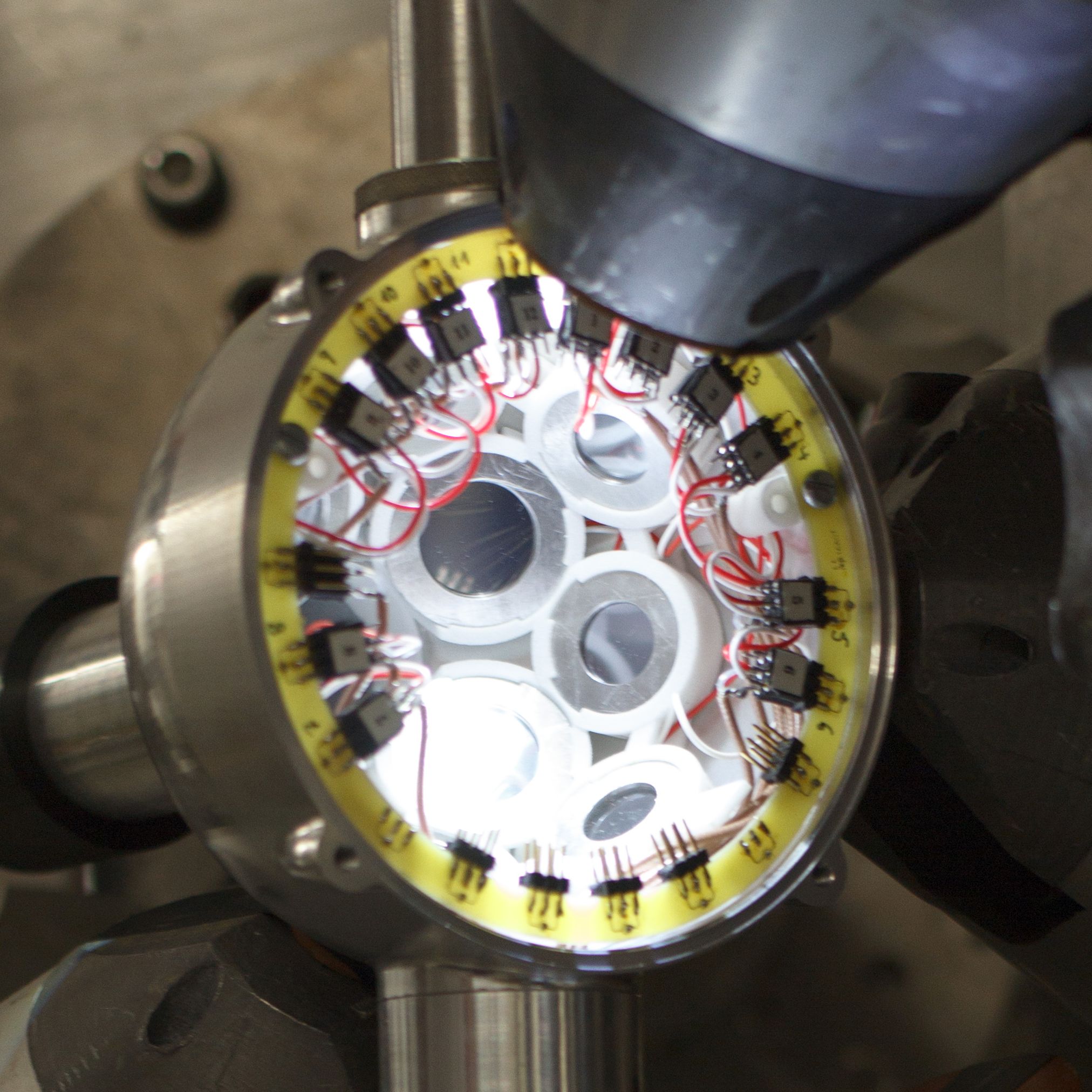}
\caption{\label{f:targetchamber} Example for using CAD geometry in \textsc{Geant4}. The original highly-detailed CAD file (t.l.) is reduced to is main components (t.r.) and converted to an STL Mesh. CADMesh then loads this mesh, which can then be assigned a material and placed like a regular solid in \textsc{Geant4} (b.l.). This process can recreate the real-life geometry (b.r.) quickly and accurately.}
\end{figure}

First, we use FreeCAD to reduce the complexity by deleting minor components that have little to no impact on the efficiency.
This should provide both a smoother mesh conversion as well as a faster simulation.
To asses which components can be deleted, we reasoned that objects that are not in the direct path between source and detector are less critical, for example the connectors in the footer of SONIC, see \cref{f:targetchamber}.
In addition, objects that are either tiny (screws) or made from low-Z material (gaskets, isolation) are also expendable in our case.
This might not hold when investigating the efficiency at very low γ-ray energies or in the X-ray regime, or scenarios where charged particles pass through.
Ideally, one could even remove the screw holes entirely, which would both be closer to reality in terms of material budget and a less complex model.

Second, we group objects made from the same material, e.g., aluminum, together and save them in a single STEP file.
Third, the STEP geometry is converted to an STL mesh.
While FreeCAD can perform this conversion, we experienced several problems, mostly stuck tracks during simulation, using this process.
Instead, we used the online STEP-to-STL converter of a 3D-print-on-demand service without issues.
An honorable mention at this point is the \emph{MeshLab} software for mesh processing and editing.
Once CADMesh loads the STL shape as tessellated volume, it can be assigned its material and placed like any other shape.
An example of this process is shown in \cref{f:targetchamber}.

\subsubsection{Detector geometry}\label{c:hpgegeo}
Several types of detectors are implemented in G4Horus, which are derived from a common \texttt{Detector} class.
This base class provides basic operations to be placeable by the \texttt{Setup} class, such that they can be mounted appropriately, see \cref{c:setup}.
\texttt{PIPS} particle detectors directly derive from this base class.

For HPGe detectors, several different crystal types exist.
A common \texttt{HPGe} base class provides implementation of the cylindrical aluminum hull, while the derived \texttt{HPGe\-Coaxial}, \texttt{HPGeClover}, and \texttt{HPGeHexagonal} classes implement the respective inner structures.
Initial parameters for most HPGe detectors were taken from the manufacturer data sheets and gathered in \texttt{DetectorLibrary}, a factory class that instantiates the correct detector from its identifier.

While all our HPGe detectors used here are technically coaxial detectors, the \texttt{HPGeCoaxial} implements the unaltered detector shape, a cylinder with a drilled hole from the back.
Data sheets provided by the manufacturer are reasonably detailed and include diameter, length, volume and distance to the end cap.
Educated guesses had to be made sometimes for the dimensions of the hole drilled for the cooling finger.

The crystals implemented by \texttt{HPGeHexagonal} are cut to semi-hexagonal conical shapes and encapsulated in hermetically closed aluminum cans of the same form \cite{Thomas1995}.
This type is used also in EUROBALL \cite{Simpson1997} and it is the predecessor to the six-fold segmented encapsulated MINIBALL \cite{Warr2013} and 36-fold segmented AGATA \cite{Akkoyun2012} detectors.
The dimensions of each crystal are identical apart from the length, which can vary slightly and is noted in the data sheets.

The implementation was tested with \isotope[226]{Ra}, \isotope[56]{Co}, and \isotope[27]{Al}(p,γ)\isotope[28]{Si} calibration data \cite{Mayer2016}.
In addition, a calibration data set with \isotope[226]{Ra}, \isotope[56]{Co}, \isotope[66]{Ga}, and \isotope[24]{Al} was used from an experiment with the SONIC-V3-ΔEE target chamber.

For most classic coaxial detectors, only minor changes, e.g., to the dead layer thickness, were necessary to reproduce the absolute FEP efficiency.
While we tried to bring the efficiency shape in line over the whole energy range, we focused less on the low energy part than described in, e.g., \cite{Chuong2016}.

Some of the encapsulated, hexagonal detectors show an experimental efficiency which is up to \SI{30}{\percent} lower as expected from simulations.
We have investigated this issue in more detail and studied the impact on the simulation accuracy at high energies, see \cref{c:geocoax}.

BGO shields for active Compton suppression were implemented with two different types of cone-shaped, lead front pieces (\emph{noses}).
Energy deposited in these detectors is converted to a veto signal afterwards.
For determining the HPGe FEP efficiency, it is not required to record veto detector data, and they can be used passively.

The two HPGe Clover detectors of the Cologne Clover Counting Setup \cite{Scholz2014a} with four crystals each were implemented with dimensions from prior work.

\subsubsection{Setup geometry}\label{c:setup}
For our experiments, detectors are placed around the target in the center.
The base class \texttt{Setup} is the abstract concept of an experimental setup which provides the common detector placement logic.
The individual setups derive from this base class and provide the Θ and φ coordinates of the mounting points as well as physical structures, if needed.

\begin{figure}
\includegraphics[width=\columnwidth]{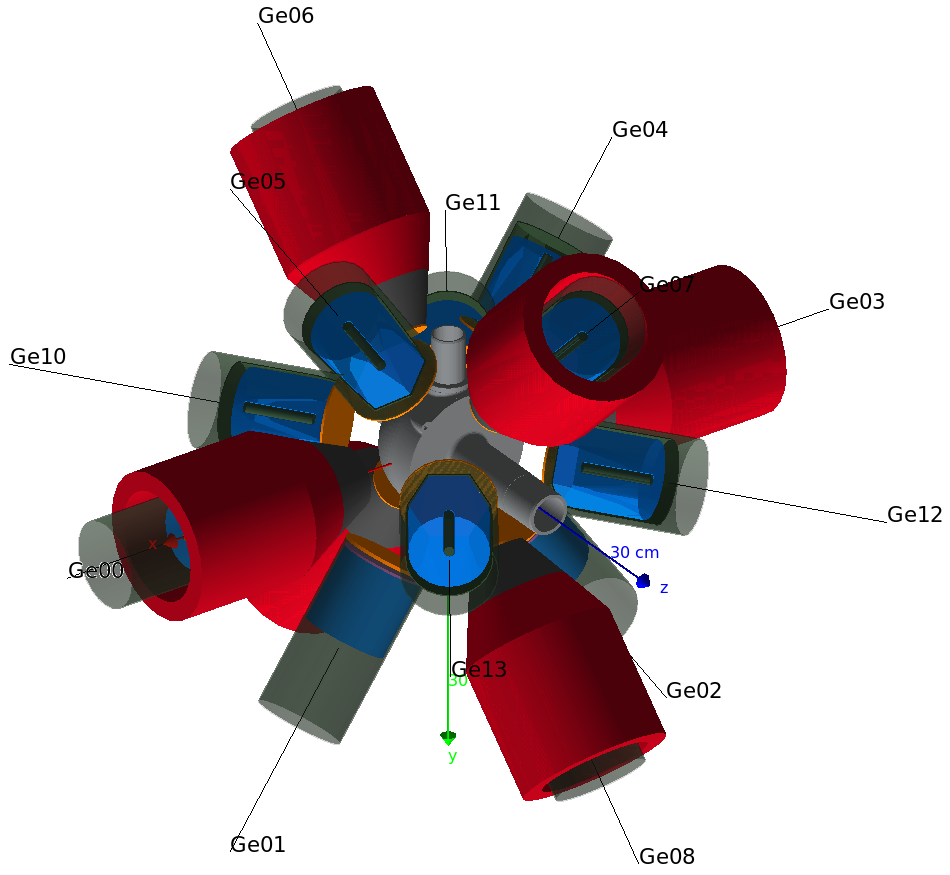}
\caption{\label{f:horus} Full virtual assembly of SONIC@HORUS. 14 HPGe detectors (blue germanium crystals with transparent black aluminum enclosures) and 6 BGO anti-Compton shields (red, with black lead noses) pointed at the target chamber (grey). Note that the z-axis points in beam direction, and the y-axis points down. Copper filters (orange) are installed in front of the detectors to reduce the number of low-energy γ-rays hitting the detectors.}
\end{figure}

The main experimental setup covered in this project is the high-efficiency γ-ray spectrometer HORUS \cite{Netterdon2014a}.
It provides 14 mounting points, labeled \texttt{Ge00} to \texttt{Ge13}, for HPGe detectors and BGO anti-Compton shields, see \cref{f:horus}.

In the center of HORUS, different target chambers can be installed.
Two different target chambers for nuclear astrophysics were implemented, one with conventional and one with CAD geometry.
Different versions of the SONIC target chamber are available via CAD geometry.
The SONIC-V3 target chamber has 12 mounting points for PIPS detectors, and its ΔE-E variant additional 12 positions to accommodate thinner PIPS detectors to form ΔE-E telescopes \cite{Pickstone2017}.

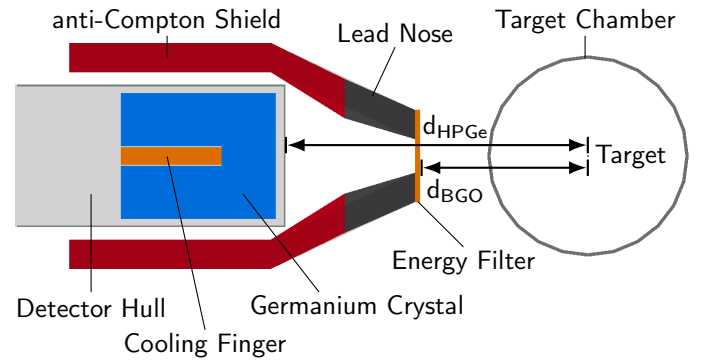
\begin{figure}
\begin{tikzpicture}[>=Latex, font=\sffamily, scale=\columnwidth/252.0pt]
\node [anchor=north west,inner sep=0] (img) at (0,-0.5) {\includegraphics[width=\columnwidth]{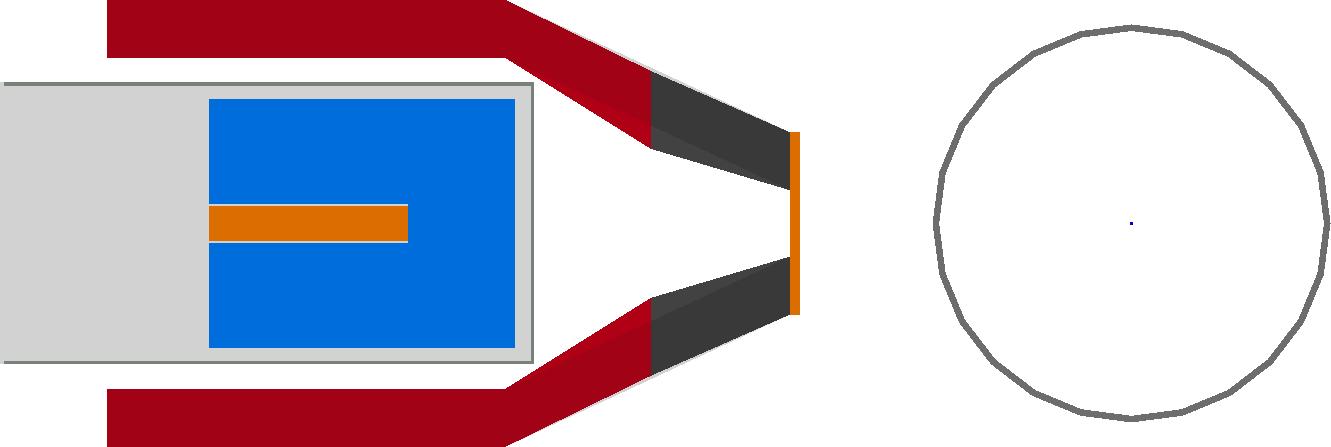}};
\node at (2,-0.2) {anti-Compton Shield};
\draw [-] (2,-0.4) -- (2,-0.7);
\node at (5,-0.35) {Lead Nose};
\draw [-] (5,-0.55) -- (4.7,-1.2);
\node at (7.5,-0.2) {Target Chamber};
\draw [-] (7.5,-0.4) -- (7.5,-0.7);
\node at (1,-4) {Detector Hull};
\draw [-] (1,-3.8) -- (1,-2.5);
\node at (4.5,-4) {Germanium Crystal};
\draw [-] (4,-3.8) -- (3,-2.5);
\node at (2.5,-4.5) {Cooling Finger};
\draw [-] (2.5,-4.3) -- (2,-2);
\node at (5.9,-3.4) {Energy Filter};
\draw [-] (5.9,-3.2) -- (5.3,-2.6);
\node at (8.1,-2) {Target};

\node at (5.8,-1.6) {d\textsubscript{HPGe}};
\draw [thick, |<->|] (7.55,-1.85) -- (3.55,-1.85);
\draw [thick, |<->|] (7.55,-2.15) -- (5.335,-2.15);
\node at (5.8,-2.45) {d\textsubscript{BGO}};
\end{tikzpicture}
\caption{\label{f:section} Schematic view of a HPGe detector and its anti-Compton shield. The distances $d_\text{HPGe}$ and $d_\text{BGO}$ are measured from the target position to the front of the detector or shield with filters equipped. For the anti-Compton shields, different nose sizes are available to match the opening angle at different distances.}
\end{figure}

For each experiment, the user builds the geometry in \texttt{DetectorConstruction} using \texttt{PlaceDetector(id, po\-si\-tion, distance, filters)}.
Within a single line, a detector is identified by its id, mounted to a named position, and equipped with passive filter materials.
See \cref{f:section} for a schematic view and distance definition.
The whole process of creating all required geometry information is thus reduced to a handful of clearly arranged lines of code, and can be done within minutes:

\begin{verbatim}
auto horus = new Horus(worldLV);
horus->PlaceDetector(
    "609502", "Ge00",  7. * cm,
    {{"G4_Cu", 2. * mm}, {"G4_Pb", 1. * mm}}
);
horus->PlaceDetector("73954",  "Ge01",  7. * cm);
// ...

auto sonic = new SonicV3(worldLV);
sonic->PlaceDetector("PIPS", "Si00", 45.25 * mm);
sonic->PlaceDetector("PIPS", "Si01", 45.25 * mm);
// ...
\end{verbatim}

This method requires recompilation on any geometry change.
While it is possible to build a messenger system to set up the geometry at runtime with \textsc{Geant4} macros, the resulting improvement in usability is currently not deemed worth the loss of direct control and flexibility.
This is a subjective matter and we might revisit this decision in the future.

\subsection{Physics}
Interactions of γ-rays are known well enough for most simulation purposes between \SI{20}{keV} and \SI{20}{\MeV}.
A predefined physics lists can supply all these interactions without hassle.
It is not necessary to create the physics from smallest components.
Most physics lists use the same standard electromagnetic physics, which, given the geometrical uncertainties, should be sufficient for this use case --- there should be no advantage in using the specialized high precision models for X-rays and low energy γ-rays.
G4Horus uses the \texttt{Shielding} physics list by default, because it includes the radioactive decay database.

\subsection{Actions}
All actions are initially dispatched by the \texttt{Action\-Ini\-tial\-iza\-tion} management class.
It parses the parameters passed to the executable and selects the appropriate primary generator, run action, and event action class.

Primary particles can either be generated by the basic \textsc{Geant4} \texttt{ParticleGun} to generate single, mono-energetic γ-rays for efficiency simulation or by specialized generators for, e.g., pγ-reactions.

One out of three output formats can be selected:
The simplest output type are histograms, which are created with the ROOT-compatible classes from \textsc{Geant4} and filled with the deposited energy for each detector.
If coincidence data is required, \texttt{ntuples} can be used.
Here, a table-like structure with a row for each detector is filled with a column for each event, also implemented with the ROOT-compatible classes from \textsc{Geant4}.
For simple efficiency simulations, this is extraordinarily inefficient as almost all entries will be zero.
Even with compression and zero-suppression, several gigabytes of data are accumulated quickly.

Instead, \emph{binary event files} can be used to store events.
They are normally produced by the sorting code \emph{SOCOv2} \cite{SOCOv2} as an intermediate event storage from raw experimental data.
Its data types, an output management class, and the respective actions have been implemented in G4Horus.
The format is well suited for the sparse data produced here, and a full simulation will produce only a few hundred megabytes of data.
The simulated data can be analyzed with the same procedure as real experimental data with the same or similar workflows.

All components are built with multi-threading in mind.
The main servers at the Institute for Nuclear Physics in Cologne provide 32 or 56 logical cores each, which can be used to capacity with the simulations.

The executable can either run in visual mode, where the geometry can be examined in 3D, or batch mode for the actual simulation.

\subsection{Automated data evaluation}
The main mission is the reliable and robust efficiency determination, which extends to simulation evaluation.
For this, a ROOT-script is included to automatically extract full-energy, single-escape, and double-escape-peak efficiencies for all simulated energies.
As energy resolution is neither included in the Monte-Carlo simulation itself nor currently added in the application, the full energy peak is a single isolated bin in the spectrum.
For single- and double-escape-peak, the Compton background is subtracted.
In case the single- and double-escape peak efficiencies must be determined with active Compton suppression, the vetoed spectra are created from \texttt{ntuple} data first.

\section{Dead regions and possible aging effects}\label{c:geocoax}
During extensive simulations of several experiments, it was found that for several hexagonally cut N-type HPGe crystals, the simulated efficiency is higher than the actual measured efficiency, up to \SI{30}{\percent} in some cases.

This issue was investigated further.
The shape of the crystal cannot be the issue, as its dimensions and especially its weight are well documented.
The dead-layer at the front of the detector was also excluded, as matching the required efficiency reduction leads to unrealistic thicknesses of \SI{10}{\mm} (instead of \SI{0.3}{\micro\m}) as well as strong deviations in the shape of the efficiency curve.

As the detectors in question were built over 20 years ago, aging effects might play a role.
The detector was used and stored cooled for most of the time but heated many times to anneal neutron induced damage.
While the dead layer at the front is created due to boron doping and should be immobile, the lithium doping of the core may have diffused further into the detector over time, creating a dead layer around the cooling finger.

Other groups have reported deviations from the manufacturer's crystal dimension specifications and aging effects.
For example, Berndt and Mortreau discovered that their cooling finger diameter is \SI{14}{\mm} instead of the declared \SI{10}{mm} by scanning the detector with highly collimated sources \cite{Berndt2012}.
Huy \emph{et al.} could trace an efficiency reduction back to an increase in the lithium dead layer of their p-type coaxial detector \cite{Huy2007}.
See also \cite{Sarangapani2017, Boson2008} and references therein.

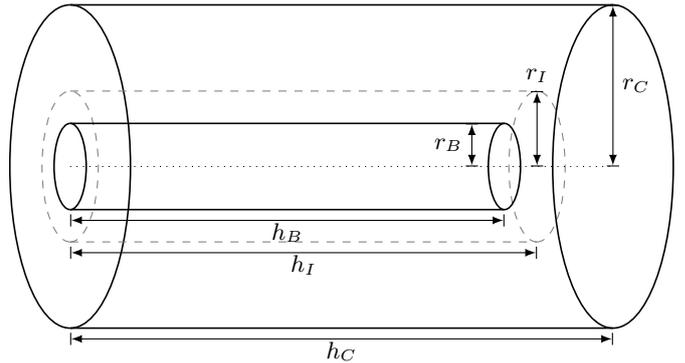
\begin{figure}
\begin{tikzpicture}[font=\small, scale=\columnwidth/6.2cm]
\tikzset{>=latex}
\draw[semithick] (0,-1.5) arc (-90:90:1.5/2.7 and 1.5);
\draw[semithick] (0,-1.5) arc (270:90:1.5/2.7 and 1.5);
\draw[semithick] (5,-1.5) arc (-90:90:1.5/2.7 and 1.5);
\draw[semithick] (5,-1.5) arc (270:90:1.5/2.7 and 1.5);
\draw[semithick] (0,-1.5) -- (5,-1.5);
\draw[semithick] (0,+1.5) -- (5,+1.5);
\draw[|<->|,thin] (0,-1.6) -- (5,-1.6) node [midway, below, yshift=0.9mm] {$h_C$};

\draw[dashed,color=gray] (0,-0.7) arc (-90:90:0.7/2.7 and 0.7);
\draw[dashed,color=gray] (0,-0.7) arc (270:90:0.7/2.7 and 0.7);
\draw[dashed,color=gray] (4.3,-0.7) arc (-90:90:0.7/2.7 and 0.7);
\draw[dashed,color=gray] (4.3,-0.7) arc (270:90:0.7/2.7 and 0.7);
\draw[dashed,color=gray] (0,-0.7) -- (4.3,-0.7);
\draw[dashed,color=gray] (0,+0.7) -- (4.3,+0.7);
\draw[|<->|,thin] (0,-0.8) -- (4.3,-0.8) node [midway, below, yshift=0.9mm] {$h_I$};

\draw[semithick] (0,-0.4) arc (-90:90:0.4/2.7 and 0.4);
\draw[semithick] (0,-0.4) arc (270:90:0.4/2.7 and 0.4);
\draw[semithick] (4,-0.4) arc (-90:90:0.4/2.7 and 0.4);
\draw[semithick] (4,-0.4) arc (270:90:0.4/2.7 and 0.4);
\draw[semithick] (0,-0.4) -- (4,-0.4);
\draw[semithick] (0,+0.4) -- (4,+0.4);
\draw[|<->|,thin] (0,-0.5) -- (4,-0.5) node [midway, below, yshift=0.9mm] {$h_B$};

\draw[dotted] (0,0) -- (5,0);
\draw[|<->|,thin] (5,0) -- (5,1.5) node [midway, right] {$r_C$};
\draw[|<->|,thin] (4.3,0) -- (4.3,0.7) node [at end, above] {$r_I$};
\draw[|<->|,thin] (3.7,0) -- (3.7,0.4) node [midway, left] {$r_B$};
\end{tikzpicture}
\caption{\label{f:deadhex-sketch}
Sketch of a HPGe crystal with radius $r_C$ and height $h_{C}$ with its borehole with radius $r_{B}$ and height $h_{B}$.
Around this hole, we assume an inactive zone with radius $r_{I}$ and height $h_{I}$.
}
\end{figure}

We simulate a possible dead layer increase by splitting the geometry of the hexagonal cut HPGe crystal (radius $r_{C}$ and height $h_{C}$) in an active and inactive part.
Here, we made the simplest possible assumption: A cylinder with radius $r_{I}$ and height $h_{I}$ around the cylindrical borehole with
radius $r_{B}$ and height $h_{B}$, see \cref{f:deadhex-sketch}.

\begin{figure}
\includegraphics[width=\columnwidth]{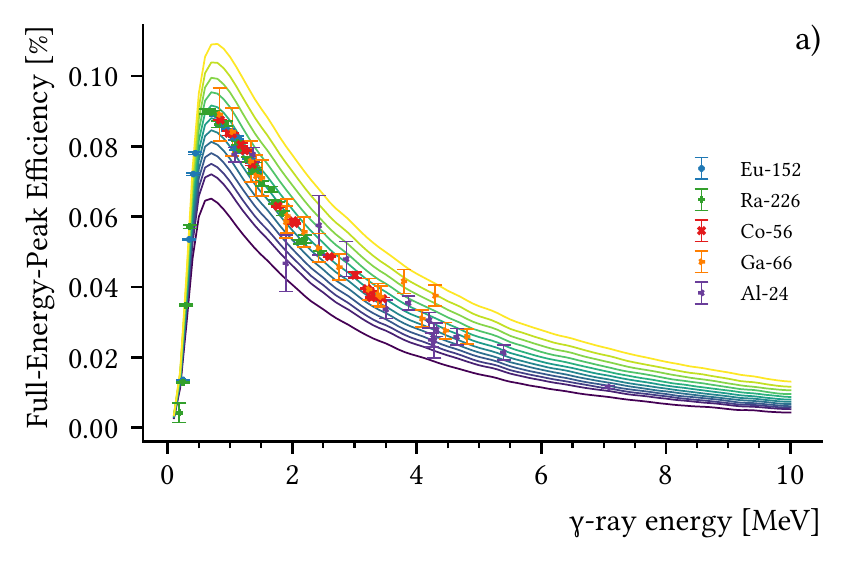}
\includegraphics[width=\columnwidth]{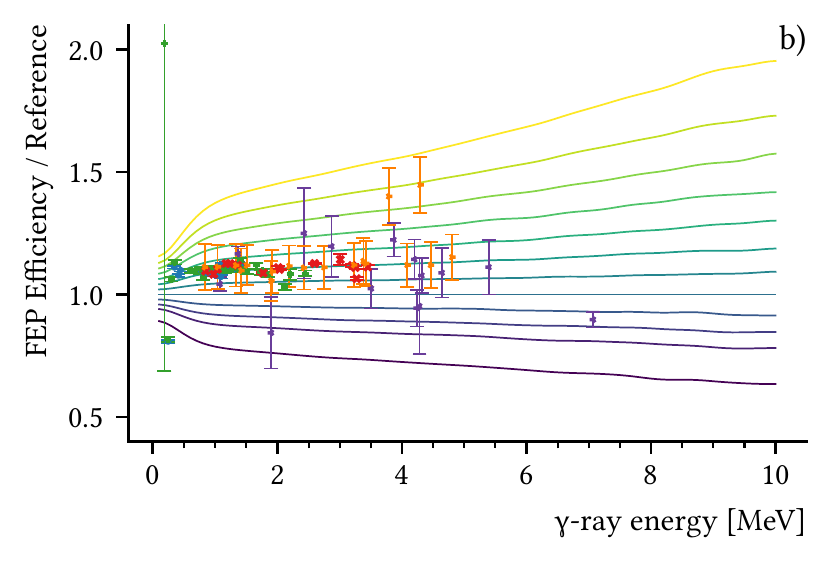}
\includegraphics[width=\columnwidth]{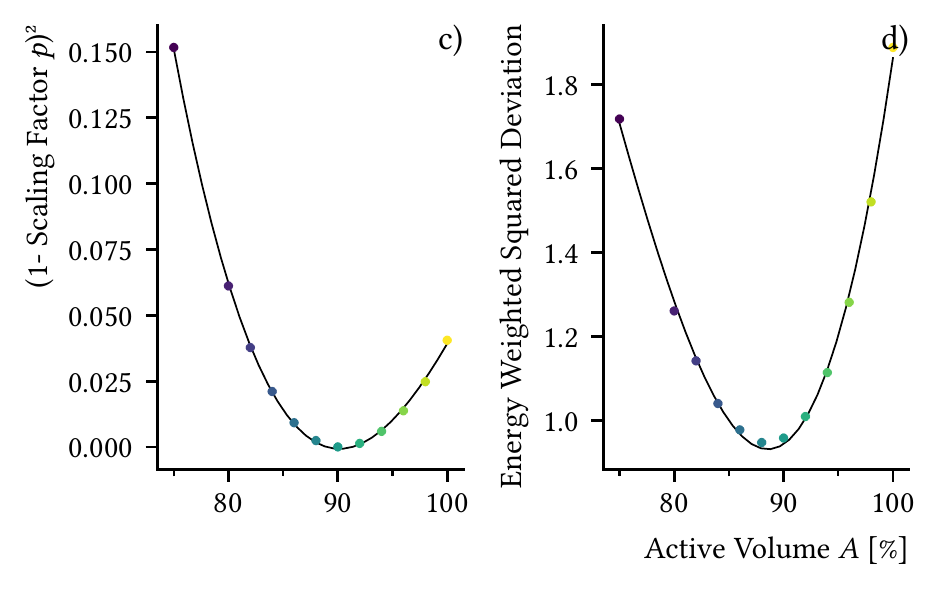}
\caption{\label{f:deadhex-efficiency}
a) Experimental and simulated full-energy-peak efficiency for a hexagonally cut encapsulated HPGe detector.
b) Experimental and simulated full-energy-peak efficiency divided by a reference simulation ($A=\SI{85}{\percent}$).
c) Scale and d) shape quality indicators for different values of active volume $A$.
Notice how the simulation for $A=\SI{100}{\percent}$ is overestimating the real performance by a significant amount - simply scaling an efficiency to the experimental values will not result in accurate results.
The relative differences between the simulations also increase drastically with γ-ray energy.
Once all geometry parameters are optimized, the minima for SCAL and EWSD should be at the same position.
}
\end{figure}

A quick approximation for $r_{I}$ and $h_{I}$ as a function of the relative active volume $A=\frac{\text{Active Volume}}{\text{Total Volume}}$ can be made in two steps:
First, the back part with the bore hole, i.e., three cylinders with the same height
\begin{equation}
A = \frac{r_C^2 - r_{I}^2}{r_C^2 - r_B^2} \Rightarrow r_{I} = \sqrt{r_C^2-A(r_C^2-r_B^2)},
\end{equation}
where a normal cylindrical shape $C$ for the whole crystal is assumed.
Second, the front part:
\begin{equation}
A = 1 - \frac{(h_{I}-h_B)r_{I}^2}{(h_C-h_B)r_C^2} \Rightarrow h_{I} = h_B + (1-A)(h_C-h_B)\frac{r_C^2}{r_{I}^2}
\end{equation}
Simulations exploring a large range of $A$ are compared to experimental values for one detector in \cref{f:deadhex-efficiency}.

The simulation should reproduce the scale and shape of the efficiency curve.
\texttt{curve\_fit} from the scipy-optimize library was used to find the scaling factor $p$ for each simulation to the measured data points.
Values between the \SI{100}{\keV}-spaced simulation points were interpolated linearly.
An ideal value would be $p=1$, i.e., no scaling. To derive the best value for $A$, this can be reformulated as a smooth minimizable function
\begin{equation}
\text{SCAL}(A) = (1-p)^2.
\end{equation}

In addition, the shape of the curve is extraordinarily important, especially with respect to deriving efficiencies at \SI{10}{\MeV}.
To emphasize the fewer calibration points at high energies, we define the energy weighted squared deviation of the scaled curve
\begin{equation}
\text{EWSD}(A) = \frac{\sum_i{E_{\gamma_i} (\epsilon_{exp}(E_{\gamma_i})-p\epsilon_{sim}(E_{\gamma_i}))^2}}{\sum_i E_{\gamma_i}},\label{eq:ewsd}
\end{equation}
which is another minimizable function of $A$ and related to the covariance / uncertainty of the scaling factor.
Note that other scaling factors for the energy could also be used, e.g., $E_{\gamma_i}^3$.
With this approach the single free variable $A$ can be determined by minimizing both SCAL and EWSD, see \cref{f:deadhex-efficiency}.

\section{Results}
\begin{figure}
\includegraphics[width=\columnwidth]{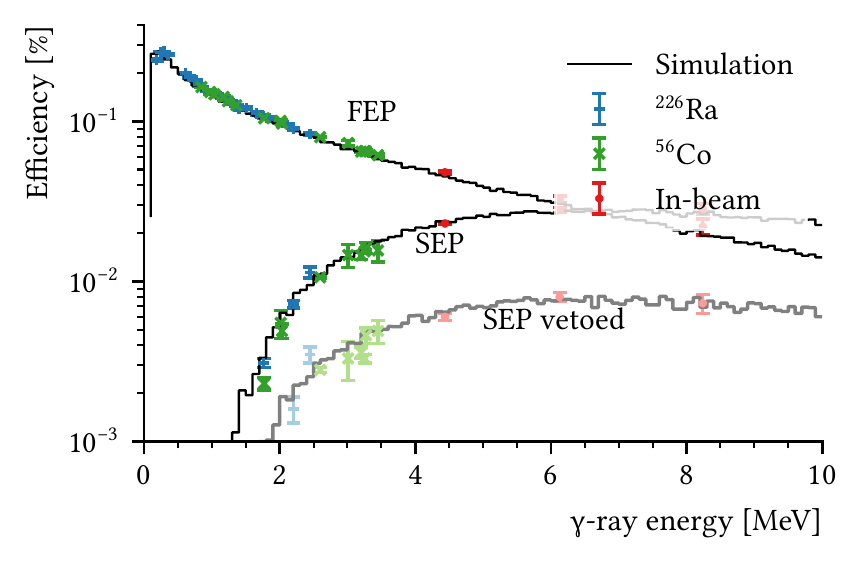}
\caption{\label{f:efficiency}
Example for simulated single escape efficiencies with and without active Compton suppression \cite{Mayer2016}. The escape-peak efficiency can also be tested in-beam with transitions from common contaminants like oxygen and carbon by scaling their intensity to the full-energy-peak efficiency.
}
\end{figure}

The goals described in \cref{c:purpose} could be achieved.
Efficiencies can be simulated with satisfactory accuracy, including SEP and DEP efficiencies with and without veto, an example is show in \cref{f:efficiency}.
In version 1.0 \cite{jan_mayer_2020_3692475}, 22 HPGe detectors and 5 target chambers are implemented and can be easily combined to the individual setup with minimal knowledge of \textsc{Geant4} or HPGe detector geometries.
Adding new or tweaking existing detectors is possible with a central data file.
There is a procedure in place to add new experimental setups and target chambers as well as detector types.
We have used this simulation environment to make informed decisions about extensions to the existing setup, e.g., adding passive shielding to reduce the number of low energetic γ-rays.

The software has been used for several experiments with good results, even though some detectors still require manual parameter tweaking to reproduce the experimental values accurately.

This project was released as Open-Source and is available from \url{https://github.com/janmayer/G4Horus} \cite{jan_mayer_2020_3692475}.
We invite everyone to adapt the project or scrounge parts of the code for other projects.

While our developments are focused on the HORUS setup, the code can be used for other, unrelated experiments employing γ-ray detectors surrounding a target.
Experimental setups can be added by deriving them from the \texttt{Setup} class and specifying the detector Θ and φ angles in the constructor.
Typical HPGe detectors can be added by appending their individual parameter sets to the Detector Library.
If the existing detector templates are insufficient, more can be added by deriving them from the \texttt{Detector} class and overriding
the provided virtual methods.
Target chamber can be implemented with the usual \textsc{Geant4} methods or with CAD-based models as described in \cref{c:chambergeo}.

\section{Outlook}
A large problem with \textsc{Geant4} is the geometry implementation.
While using code is a step up over digital punch cards used in MCNP, it is decades behind other simulation integrations as seen in, e.g., finite element analysis.
In the future, it would be advisable to find a modern solution that is ready for everyday production usage.
Due to its massive advantages in development speed, ease of use, and flexibility, CAD based simulation geometry could be officially supported by the \textsc{Geant4} collaboration.
To reduce the slowdown of simulations, a hybrid approach might be feasible: Convert structures to simple shapes where possible and use tessellated shapes for the remnants.
In a new Monte Carlo code, only tessellated shapes could be supported and used exclusively with GPUs.

For G4Horus, we continue to make improvements to the description of our detectors as well as add new functionality like better support for pγ- and γγ-coincidence measurements.

\section{Acknowledgments}
We would like to thank D. Diefenbach and S. Thiel from our development workshop for accelerators and accelerator experiments for designing the target chambers and their help with the CAD models, Dr. J. Eberth for the fruitful discussions about HPGe detectors, and C. Müller-Gatermann and the accelerator crew for their help with the experiments and source production.

Supported by the DFG (ZI 510/8-1, ZI-510/9-1).

\bibliography{g4horus}

\end{document}